\documentclass[epj,final]{svjour}

\usepackage{graphicx}
\usepackage{dcolumn}   
\usepackage{bm}        
\usepackage{amssymb}   
\usepackage{textcomp}
\usepackage{ucs}
\usepackage{hyperref}
\usepackage{amsmath}
\usepackage{color}
\usepackage{footnote}

\begin{document}

\title{Attenuation of vacuum ultraviolet light in pure and xenon-doped liquid argon - an approach to an assignment of the near-infrared emission from the mixture}

\author{A. Neumeier\inst{1} \and T. Dandl\inst{2} \and A. Himpsl\inst{2} \and L. Oberauer\inst{1} \and W. Potzel\inst{1} \and S. Sch\"onert\inst{1} \and A. Ulrich\inst{2}\thanks{\emph{Andreas Ulrich:} andreas.ulrich@ph.tum.de 
}
}                     
\institute{Technische Universit\"at M\"unchen, Physik-Department E15, James-Franck-Str. 1, D-85748 Garching, Germany \and Technische Universit\"at M\"unchen, Physik-Department E12, James-Franck-Str. 1, D-85748 Garching, Germany}

\date{Published in EPL (2015)}

\abstract{
Results of transmission experiments of vacuum ultraviolet light through a 11.6\,cm long cell filled with pure and xenon-doped liquid argon are described. Pure liquid argon shows no attenuation down to the experimental short-wavelength cut-off at 118\,nm. Based on a conservative approach, a lower limit of 1.10\,m for the attenuation length of its own scintillation light could be derived. Adding xenon to liquid argon at concentrations on the order of parts per million leads to strong xenon-related absorption features which are used for a tentative assignment of the recently found near-infrared emission observed in electron-beam excited liquid argon-xenon mixtures. Two of the three absorption features can be explained by perturbed xenon transitions and the third one by a trapped exciton (Wannier-Mott) impurity state. A calibration curve connecting the equivalent width of the absorption line at 140\,nm with xenon concentration is provided.
}  

\maketitle

\section{Introduction}

In an earlier study of the attenuation of vacuum ultraviolet (VUV) light \cite{neumeier_epjc} we had found that xenon impurities lead to strong absorption features in the region of the resonance lines of xenon. The first observed absorption feature appears at a wavelength of 126.5\,nm and can be attributed to a trapped-exciton impurity-state (Wannier-Mott type) whereas a second absorption feature at 141.0\,nm  can be attributed to a perturbed atomic transition of xenon ($^{1}S_{0} \rightarrow$ $^{3}P_{1}$, 146.96\,nm) \cite{jortner}. Distillation was used to reduce the strength of these absorption features. However, a reduced transmission towards short wavelengths remained (Fig. 6, upper panel in ref.\,\cite{neumeier_epjc}). It was not clear whether this was due to a remaining impurity or due to absorption or scattering of light in liquid argon itself. In the present publication we show that the reduced transmission was due to a xenon impurity and that an improved distillation technique leads to {an essentially} transparent liquid argon sample {of 11.6\,cm length}. The technical details of the transmission measurements presented here are published in ref.\,\cite{neumeier_jinst}.

Adding xenon on a level of parts per million deliberately to the practically perfectly purified liquid argon leads to absorption features of increasing strengths with increasing xenon concentrations as described below. This study is part of a series of measurements of the particle-beam induced scintillation as well as the attenuation of scintillation light in liquid noble gases \cite{neumeier_epjc,neumeier_jinst,heindl_epl,heindl_jinst,neumeier_epl,hofmann_epjc,neumeier_epl_2}. Liquid noble gases are planned to be used or are already being used in very large scintillation detectors. Experiments in the sector of rare event physics like the direct dark matter search \cite{ArDM,DarkSide,MiniCLEAN,DEAP_1,DEAP_2,DARWIN} and high-energy neutrino physics \cite{ICARUS,LBNE} make use of liquid argon as the detector material. Characterizing the optical properties of this material will help in analysing the data and designing new detector concepts. {Here we will only present the transmission data for pure and xenon-doped liquid argon.}
Although the concept for measuring the attenuation of light in a liquid noble-gas sample is straightforward we identified various, partly surprising sources of systematic effects which influence the results. These effects are discussed in detail in a separate publication focused on technical aspects \cite{neumeier_jinst}. The systematic uncertainties from those effects are included in the analysis of the data presented below.

\section{Experimental Setup}

The experimental setup is similar to that in ref.\,\cite{neumeier_epjc} and described in detail in ref.\,\cite{neumeier_jinst}. Briefly, the inner cell containing the liquid noble gas mixtures has an optical path length of 11.6\,cm. The length of the inner cell was limited by the size of the vacuum cell (outer cell) surrounding the inner cell. The vacuum ultraviolet light from a deuterium arc lamp entered and exited the liquid argon-xenon mixtures through two MgF$_{2}$ windows in the optical path. A modification of the gas system in \cite{neumeier_epjc} allowed {us} to perform fractional distillation of the noble gas samples in a continuous flow mode prior to condensation into the inner cell. The liquid argon-xenon mixtures were prepared in the following way: A small amount of xenon was filled into the gas system. The quantity was determined by the pressure which was measured with a precise capacitive manometer (MKS Baratron 390H 1000). Then argon gas was added until a pressure of 1400\,mbar was reached and the mixture was purified for about 1.5\,hours and finally condensed into the cell for absorption measurements\footnote{The purification of the gas samples during the measurements was maintained by bypassing the cryogenic cell via a bypass valve and purifying the buffer volume which is connected to the cryogenic cell in a continuous flow mode.}. The condensation  was stopped at a pressure of about 1000\,mbar when the cell was completely filled with the liquid. 

However, it has to be noted that we cannot quantify exactly the concentration of xenon in liquid argon since the mixtures were prepared in the gas phase. Basically there are two counter acting effects which have to be taken into account. Firstly, the cryogenic cell acts as a cold trap for the remaining xenon in the buffer volume. Therefore an upper limit for the real concentration can be given in the extreme case that all the xenon from the gas system (buffer volume) is condensed into the cryogenic cell. For the real xenon concentration in the liquid this upper limit corresponds to a factor\footnote{After a pressure reduction in the gas system by 400\,mbar (from 1400 to 1000\,mbar) the inner cell was completely filled with the liquid. If all the xenon from the gas system would be condensed into the cold inner cell a maximum deviation of the xenon content in the liquid by a factor of $\frac{1400\,mbar}{400\,mbar}=3.5$ can be expected.} of 3.5 compared to the xenon concentration prepared in the gas phase. Secondly, a certain amount of xenon can be condensed to the walls of the cryogenic cell which leads to a reduction of the real xenon concentration in the liquid. An aggregation of xenon in liquid argon seems verly unlikely for concentrations below 30\,ppm \cite{jortner}. The systematic errors described on the actual concentrations in the liquid phase could only be reduced by dedicated measurements of the solubility of xenon in liquid argon.

\section{VUV-light attenuation in liquid argon revisited}

\begin{figure}[h]
 \centering
 \includegraphics[width=\columnwidth]{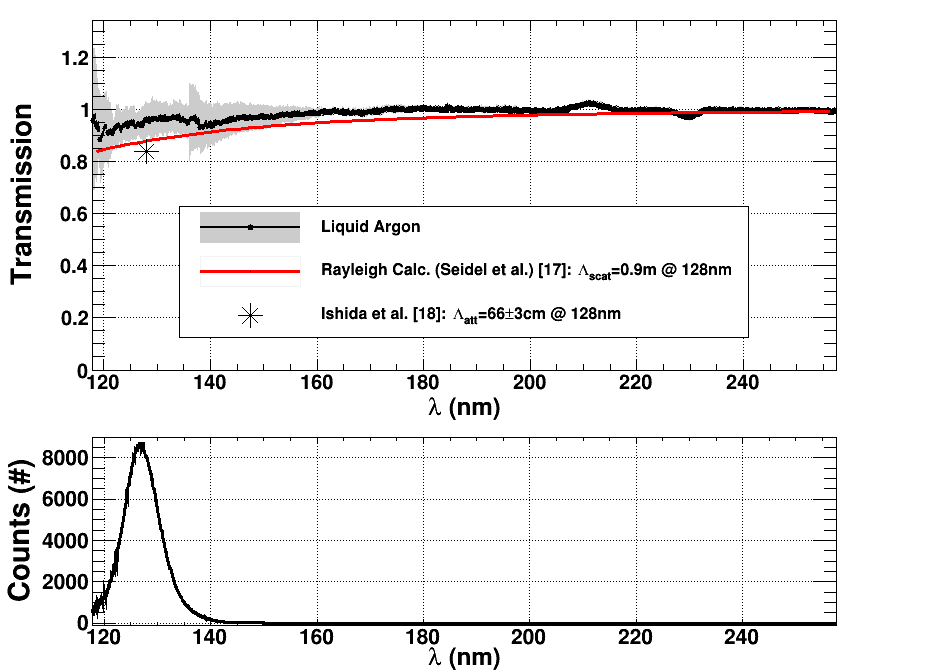}
 \caption{\textit{The measured transmission of 11.6\,cm pure liquid argon is shown {(black curve with grey error bars)}. On the current level of sensitivity no xenon-related absorption effects could be identified and pure liquid argon is fully transparent down to the short-wavelength cut-off of the system at 118\,nm. The red line shows the expected transmission of a sample of 11.6\,cm pure liquid argon according to a Rayleigh scattering length of 90\,cm at 128\,nm ($\Lambda_{scat}$ = 0.9\,m at 128\,nm) \cite{seidel_rayleigh}. The black asterisk shows, for comparison, the expected transmission value of 11.6\,cm pure liquid argon for an attenuation length of (66$\pm$3)\,cm at 128\,nm adopted from ref.\,\cite{ishida_attenuation}. The lower panel shows, for comparison, the electron-beam induced emission of pure liquid argon (adopted from ref.\,\cite{heindl_epl}). 
}}
 \label{fig:lar_emission_transmission}
\end{figure}

As mentioned in the introduction we had found a reduced transmission in liquid argon from about 130\,nm down to the wavelength limit of our apparatus of 118\,nm. The remaining transmission was about 70\,\% at 118\,nm in the 5.8\,cm long cell \cite{neumeier_epjc}. Fig.\,\ref{fig:lar_emission_transmission} (upper panel) shows the transmission of pure liquid argon with the improved distillation technique. The transmission of pure liquid argon is improved to a level where it is difficult to quantify an attenuation length with the present length (11.6\,cm) of the inner cell and the optical measuring technique used here. However, from a comparison of the lower limits of the error bars in Fig.\,\ref{fig:lar_emission_transmission} it is possible to derive a conservative lower limit of 1.10\,m for the attenuation length of pure liquid argon for its own scintillation light. This means that the reduced transmission in Fig.\,6 in ref.\,\cite{neumeier_epjc}\footnote{In Fig.\,6 (upper panel) of ref.\,\cite{neumeier_epjc} there are still two very weak absorption features visible at about 126.5\,nm and 140\,nm due to a xenon impurity.} was still due to a remaining xenon impurity.

Although we can not provide an attenuation length for pure liquid argon in the VUV we can comment on a prediction concerning Rayleigh scattering in liquid argon. In ref.\,\cite{seidel_rayleigh} it is calculated that the Rayleigh scattering length is 90\,cm at the scintillation wavelength of pure liquid argon (128\,nm). Since the Rayleigh scattering cross section scales inversely with the fourth power of wavelength, the results presented in Fig.\,\ref{fig:lar_emission_transmission} indicate a Rayleigh scattering length larger than 90\,cm at a wavelength of 128\,nm. The solid red curve in the upper panel of Fig.\,\ref{fig:lar_emission_transmission} shows the calculated wavelength-dependent transmission of pure liquid argon for a Rayleigh scattering length of 0.90\,m at 128\,nm. With a length of the optical path of 11.6\,cm and an assumed Rayleigh scattering length of 0.90\,m a transmission of 88\% could be expected at 128\,nm which seems to be too small\footnote{The lower limit of the error bar at 128\,nm in Fig.\,\ref{fig:lar_emission_transmission}, upper panel, is at 90\,\% transmission. This corresponds to a lower limit of 1.10\,m for the attenuation length.}. However, the decreasing trend of the measured transmission towards the short wavelength end of the experimental setup could be a hint towards beginning of Rayleigh scattering. {A Rayleigh scattering length of 90\,cm can presently neither be confirmed nor excluded conclusively at the moment \cite{neumeier_jinst}.} Furthermore a second, wavelength-integrated measurement of the attenuation length of liquid argon is also shown for comparison (see the asterisk in the upper panel of Fig.\,\ref{fig:lar_emission_transmission}). Ishida et al. \cite{ishida_attenuation} measured a wavelength-integrated attenuation length of $66 \pm 3$\,cm. With an optical path length of 11.6\,cm a transmission of approximately 84\,\% at 127\,nm could be expected. {The present experiment provides a lower limit of 1.10\,m for the attenuation length.}   

\section{VUV-absorption of xenon-doped liquid argon}
 
\begin{figure}[h]
 \centering
 \includegraphics[width=\columnwidth]{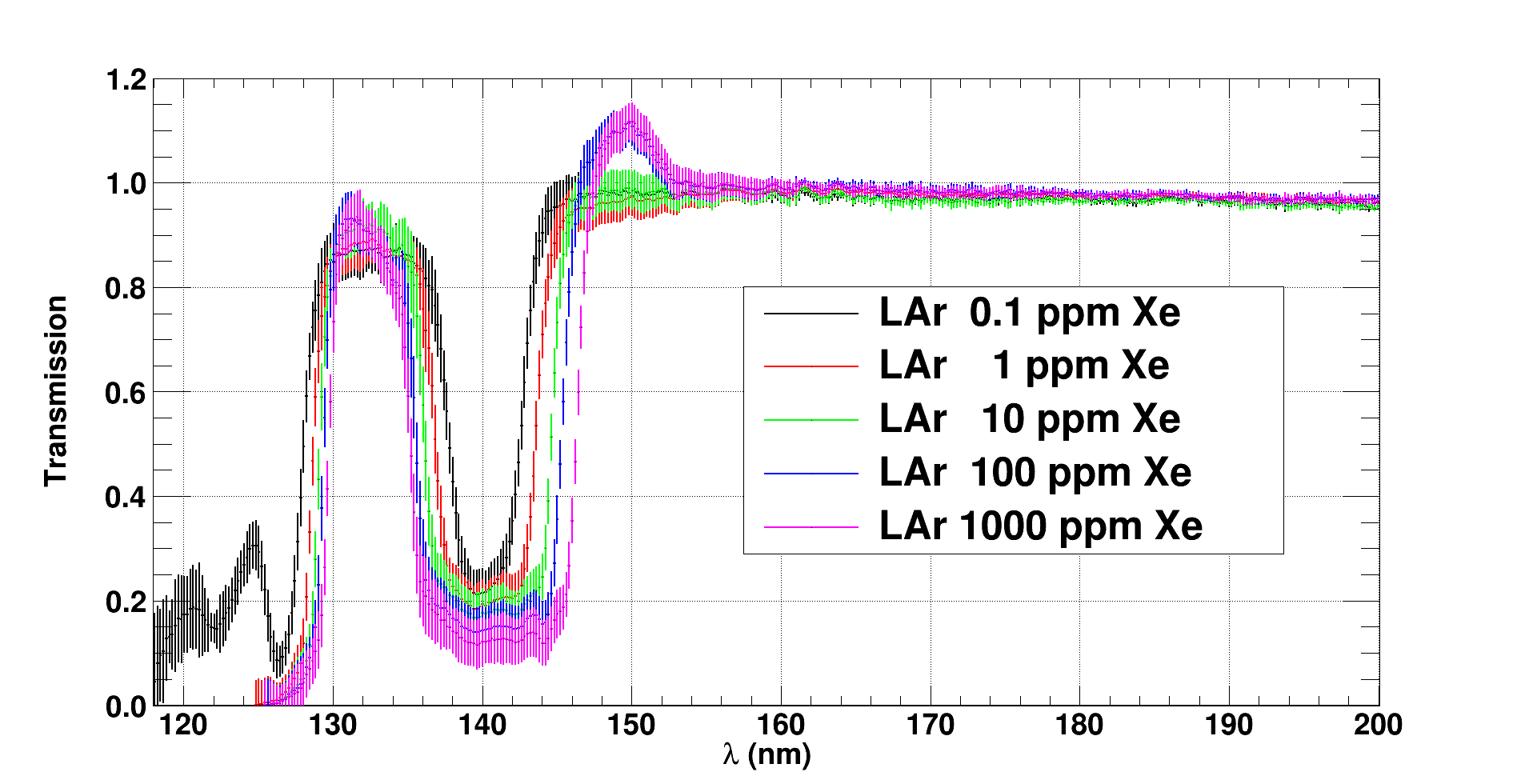}
 \caption{\textit{The transmission of 11.6\,cm liquid argon doped with different amounts of xenon is presented. The xenon concentrations have been increased by factors of 10 from 0.1 to 1000\,ppm. The most prominent absorption band at 140\,nm broadens with increasing xenon concentration. For a xenon concentration above 0.1\,ppm the mixture is not transparent below 130\,nm for a length of the optical path of 11.6\,cm. Three different absorption bands related to xenon can be identified in liquid argon doped with 0.1\,ppm xenon. The center {wavelengths} of the transmission minima are at 122.2\,nm and 140.0\,nm and can be attributed to perturbed $^{1}P_{1}$ (n=1, $^{2}P_{\frac{1}{2}}$) and $^{3}P_{1}$ (n=1, $^{2}P_{\frac{3}{2}}$) atomic xenon states \cite{jortner}. The third transmission minimum is centered at 126.5\,nm and can be attributed to a trapped exciton (n=2, $^{2}P_{\frac{3}{2}}$ Wannier-Mott) impurity state \cite{jortner}. The transmission above unity at $\sim 150$\,nm for the 100\,ppm and 1000\,ppm mixtures can be attributed to resonance fluorescence (details see text). This feature is also visible in the electron-beam induced emission spectra of these mixtures \cite{neumeier_epl_2}.}}
 \label{fig:transmission_larxe}
\end{figure}

An overview over VUV transmission spectra in xenon-doped liquid argon with xenon concentrations increased in steps by factors of 10 is shown in Fig.\,\ref{fig:transmission_larxe}. Xenon-doped liquid argon shows three strong xenon-related absorption features. The most prominent absorption band is centered at 140\,nm and broadens with increasing xenon concentration. In the transmission spectrum of liquid argon doped with 0.1\,ppm xenon three different absorption bands can be identified. The center wavelengths of the transmission minima are at 122.2 and 140.0\,nm and can be attributed to perturbed $^{1}P_{1}$ (n=1, $^{2}P_{\frac{1}{2}}$) and $^{3}P_{1}$ (n=1, $^{2}P_{\frac{3}{2}}$) atomic xenon states \cite{jortner}. The third transmission minimum is centered at 126.5\,nm and can be attributed to a trapped exciton (n=2, $^{2}P_{\frac{3}{2}}$ Wannier-Mott) impurity state \cite{jortner}. Increasing the xenon concentration leads to a broadening of the 140\,nm absorption band. The 126.5\,nm absorption band becomes completely opaque below 130\,nm for xenon concentrations higher than 0.1\,ppm and an optical path length of 11.6\,cm. 

Both, the 100 and the 1000\,ppm xenon in liquid argon mixtures show a transmission above unity with a peak wavelength at approximately 150\,nm. The mixtures investigated here in transmission have also been studied wavelength resolved in emission \cite{neumeier_epl_2} by electron-beam excitation. It turned out that there are also emission features at this wavelength which are dependent on the concentration of xenon in liquid argon. Therefore, we tentatively attribute this transmission above unity to resonance fluorescence. Either the formation of weakly bound ArXe* excimers \cite{Nowak} which lead to the weak emission feature centered at 149\,nm or a direct transition ($^{3}P_{2} \rightarrow$ $^{1}S_{0}$) of excited xenon atoms to the ground state and thereby leading to the characteristic resonance peak at 147\,nm \cite{nist} could explain this emission feature.

\begin{figure}[h]
 \centering
 \includegraphics[width=\columnwidth]{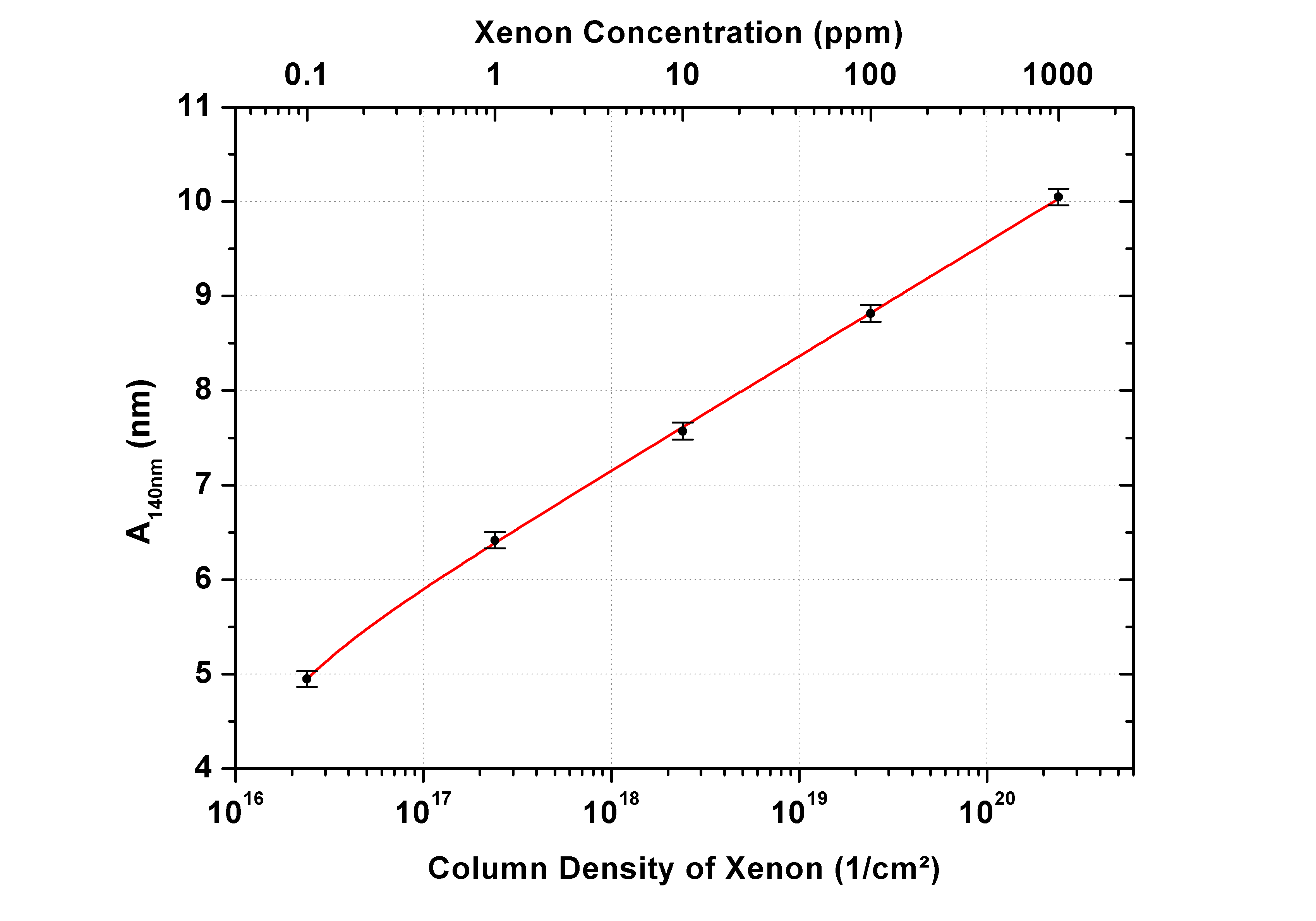}
 \caption{\textit{The equivalent widths of the 140\,nm transmission minima from Fig.\,\ref{fig:transmission_larxe} for liquid argon doped with 0.1 to 1000\,ppm xenon are presented (black dots). The integration was performed from 132\,nm to 147\,nm according to equation (\ref{eq:equivalent_width}). The y-axis shows the equivalent width $(A_{140\,nm})$ and the lower x-axis the column density of xenon atoms in
 liquid argon. The column density is the density of xenon atoms times the optical path length (here 11.6\,cm). The upper x-axis shows, for comparison, the corresponding xenon concentrations when the argon-xenon mixtures were prepared. The data points are fitted with a natural logarithmic function represented by the red line: $y=a-b \cdot \ln\left(x+c\right)$.
 The best-fit parameters have been determined as:
 $a=-14.6\pm0.4,b=-0.525\pm0.008,c=-(9\pm2)\cdot10^{15}$.}}
 \label{fig:wachstumskurve}
\end{figure}

To quantify the influence of the xenon concentration in liquid argon the area of the transmission minimum at 140\,nm has been calculated for all five measured concentrations. The result is presented in Fig.\,\ref{fig:wachstumskurve} (black dots) and has been obtained from integrating the spectra shown in Fig.\,\ref{fig:transmission_larxe} from 132\,nm to 147\,nm according to equation (\ref{eq:equivalent_width}):

    \begin{equation}
     \label{eq:equivalent_width}
     A_{140\,nm} = \int\limits_{132\,nm}^{147\,nm} (1-T(\lambda))d\lambda
    \end{equation}
    
    \begin{description}
    \item [$A_{140\,nm}$:] Area or equivalent width of the transmission minimum at 140\,nm center wavelength
    \item [$T(\lambda)$:] Measured transmission in Fig.\,\ref{fig:transmission_larxe}
    \end{description}

The equivalent width $A_{140\,nm}$ of the transmission minima scales logarithmically with xenon concentration in liquid argon and is descriptively explained as follows: The xenon-related absorption feature has a certain concentration-dependent area. The equivalent width of this absorption feature is the width which would be necessary to form a rectangular absorption feature with the same area which
drops to zero transmission.

The area of the transmission minimum at 140\,nm provides information for the determination of the xenon concentration in liquid argon. If wavelength-resolved information on the transmission of liquid argon deliberately doped (or contaminated in the case of pure liquid argon) with xenon
is available, the xenon concentration can be estimated using the calibration curve
presented in Fig.\,\ref{fig:wachstumskurve}. The x-axis in Fig.\,\ref{fig:wachstumskurve} shows the so-called column density which is the concentration of xenon atoms times the optical path length (11.6 cm in the present setup). An estimate of the column density of xenon atoms in liquid argon can be calculated according to equation (\ref{eq:column_density}):

    \begin{equation}
     \label{eq:column_density}
     x = \exp\left(\frac{a-y}{b}\right)-c \ \ or \ \ y=a-b \cdot \ln\left(x+c\right)
    \end{equation}

    To obtain the column density $x$ of xenon atoms in $\frac{1}{cm^{2}}$ the area or equivalent width $y$ of the transmission minimum at 140\,nm center wavelength (integrated from 132 to 147\,nm) has to be measured in nm. The parameters a,b and c are the best-fit values derived from the phenomenological fit of the calibration curve in Fig.\,\ref{fig:wachstumskurve}: 
    
    \begin{itemize}
    \item[] $a=-14.6\pm0.4$
    \item[] $b=-0.525\pm0.008$
    \item[] $c=-(9\pm2)\cdot10^{15}$
    \end{itemize}
    
\section{Interpretation of the transmission minima - A tentative assignment of the near-infrared emission in xenon-doped liquid argon}

It is known from experiments with pulsed electric discharges in xenon-doped (10\,ppm - 1000\,ppm) gaseous argon that xenon is a very efficient electronic energy acceptor in argon-xenon mixtures \cite{Gedanken}. In a series of experiments studying the emission spectra of electron-beam excited liquid argon-xenon mixtures, efficient energy transfer from argon to xenon was also observed in the liquid phase \cite{neumeier_epl,neumeier_epl_2}. In particular, a novel infrared emission could be detected in xenon-doped liquid argon \cite{neumeier_epl}. However, so far an assignment of the near-infrared emitting species was not possible. In a way similar to the energy transfer processes in the gas phase described in ref.\,\cite{Gedanken}, a model for the liquid phase is proposed and briefly explained\footnote{For further details, see ref.\,\cite{neumeier_phd}.}. It has to be emphasized that the following model is just an attempt for an assignment of the near-infrared emitting species. However, the model can qualitatively explain the emission wavelength, the time structure and the xenon concentration dependence of the recently found near-infrared emission. 

\begin{figure}[h!]
        \centering
        \includegraphics[width=\columnwidth]{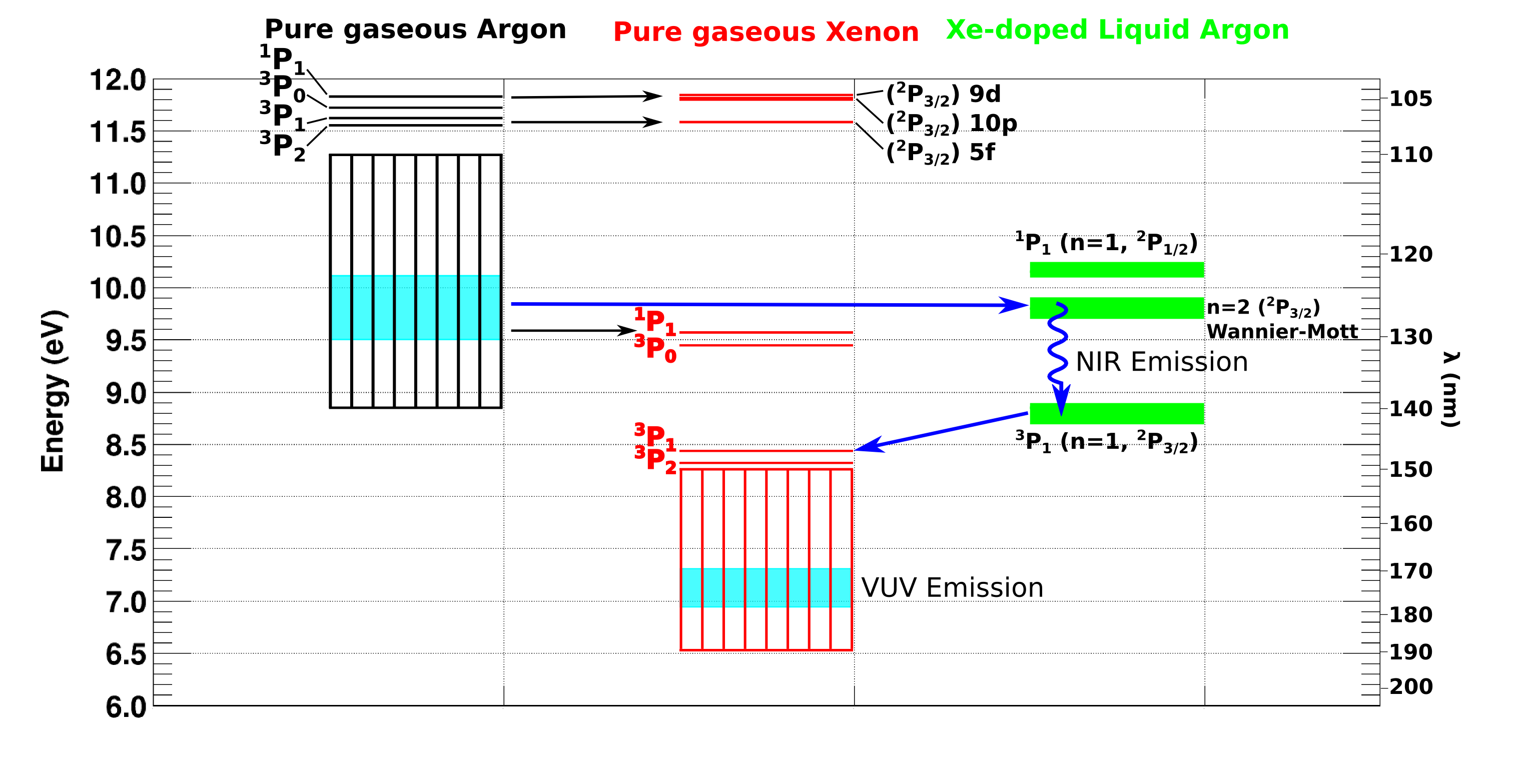}
        \caption{\textit{(Adopted from \cite{neumeier_phd}.) The lower atomic levels and the energy released in molecular emissions, respectively, of pure gaseous argon (black), pure gaseous xenon (red), and xenon-doped liquid argon (green) are shown. The atomic levels are adopted from \cite{nist}. The left ordinate shows the energy in eV and the right ordinate shows the corresponding wavelength in nm. The vertically shaded areas denote the energy regions where the second excimer continua of pure gaseous argon and pure gaseous xenon are located. The blue regions in the vertically shaded areas denote the peak emission wavelengths with full width at half maximum of the corresponding excimer emissions. The green bands correspond to the energy levels where strong xenon-related absorption features have been measured. The absorption features can be seen as transmission minima in xenon-doped liquid argon in Fig.\,\ref{fig:transmission_larxe}. The identification of the different states in xenon-doped liquid argon is adopted from \cite{jortner}. The blue arrows indicate energy transfer and emission processes which could explain the observed near-infrared and VUV emissions in xenon-doped liquid argon. The recently found near-infrared emission is tentatively assigned to a transition from a n=2 ($^{2}P_{\frac{3}{2}}$) Wannier-Mott impurity state to a perturbed $^{3}P_{1}$ (n=1, $^{2}P_{\frac{3}{2}}$) state of xenon. The thick green lines in xenon-doped liquid argon rather represent band structures in disordered systems than atomic transitions in gases.}}
        \label{fig:ArXe_Energieniveaus}
    \end{figure}

In Fig.\,\ref{fig:ArXe_Energieniveaus} the energy transfer processes from argon to xenon in the gas and in the liquid phase are depicted schematically \cite{neumeier_phd}. The energy transfer processes from argon to xenon in the gas phase are indicated by the black arrows in Fig.\,\ref{fig:ArXe_Energieniveaus} and explained very detailed in ref.\,\cite{Gedanken}. Therefore, only the important aspects will briefly be mentioned here. In argon-xenon mixtures two energy transfer-processes which have high and comparable energy transfer cross-sections lead to an efficient energy transfer from argon to xenon. This is a unique feature of argon-xenon mixtures which is absent in argon-krypton mixtures. Firstly, the close coincidence between the 9d (J=1, J=2) levels of pure xenon and the ($^{1}P_{1}$) state of pure argon leads to efficient atom-atom energy transfer (indicated by the black arrows in the upper part of Fig.\,\ref{fig:ArXe_Energieniveaus}). Secondly, due to the overlap between the energy released by the decay of argon excimers (second excimer continuum of pure argon) and the energy of the $^{1}P_{1}$ state of pure xenon also an energy transfer from argon excimers to xenon atoms in the ground state is possible which are then excited to the $^{1}P_{1}$ state.

As the density increases from the gas phase to the liquid phase, the discrete level scheme of isolated atoms is changed to band structures in a disordered insulator. This is also visible in the transmission of xenon-doped liquid argon (see Fig.\,\ref{fig:transmission_larxe}), since the transmission minima in the vacuum ultraviolet wavelength region are rather broad absorption features than discrete atomic absorption lines. The absorption feature at 140\,nm center wavelength in Fig.\,\ref{fig:transmission_larxe} has the lowest excitation energy. This absorption feature is assigned to a perturbed $^{3}P_{1}$ (n=1, $^{2}P_{\frac{3}{2}}$) atomic xenon state which is only shifted by $\sim0.5$\,eV (7\,nm blue shifted) towards higher energies compared to the undisturbed isolated xenon atom \cite{jortner}. However, the band structures above the first excited state in xenon-doped liquid argon are significantly changed compared to the isolated xenon atom due to the surrounding argon atoms. The formation of a trapped exciton (n=2, $^{2}P_{\frac{3}{2}}$, Wannier-Mott) impurity state \cite{jortner} of xenon in liquid argon is a unique feature of the liquid and solid phases and has no parentage in pure gaseous xenon.

A transition (see Fig.\,\ref{fig:ArXe_Energieniveaus}) with an energy difference of $\sim$1\,eV between the Wannier-Mott (n=2, $^{2}P_{\frac{3}{2}}$) impurity state in xenon-doped liquid argon and the perturbed $^{3}P_{1}$ (n=1, $^{2}P_{\frac{3}{2}}$) atomic xenon state could explain the emitted wavelengths from $\sim1030$\,nm to $\sim1550$\,nm which coincide almost perfectly with the measured wavelength region ($\sim1100$\,nm - $\sim1550$\,nm, see Fig.\,1, upper panel in ref.\,\cite{neumeier_epl}). At the moment no conclusive statement can be made whether further xenon atoms in the ground state are involved forming an excimer-like potential with xenon atoms in the Wannier-Mott impurity state. This indeed could explain the continuum character of the near-infrared emission. However, a comparison of the relative variation of the integrated near-infrared signal (green triangles in Fig.\,4 in ref.\,\cite{neumeier_epl_2}) for xenon concentrations of 0.1, 1 and 10\,ppm in liquid argon leads to the conclusion that the trend can be described by an initially linearly increasing signal with xenon concentration followed by a beginning saturation. A quadratically increasing signal seems unlikely since the variation of the near-infrared signal is too low. Therefore, a light-emission process where two xenon atoms are involved seems unlikely since then a quadratically increasing near-infrared signal with xenon concentration would be expected. Consequently, the band structure alone could also be responsible for the continuous character of the near-infrared emission. For the formation of the Wannier-Mott impurity states in xenon-doped liquid argon the following energy transfer reaction is proposed \cite{neumeier_phd}:
    
    \begin{small}
        \begin{equation}
        Ar_{2}\left(^{1,3}\Sigma_{u}\right) + Xe\left(^{1}S_{0}\right) \rightarrow 2Ar\left(^{1}S_{0}\right)+Xe \left(n=2,\,^{2}P_{\frac{3}{2}}\right)
        \end{equation}
    \end{small}
    
Note that the excitation energy of the Wannier-Mott impurity state \cite{jortner} coincides almost exactly with the peak emission wavelength of the excimer emission of pure argon which could be a strong indication for an efficient energy transfer from argon to xenon. This energy transfer is illustrated in Fig.\,\ref{fig:ArXe_Energieniveaus} by the blue horizontal arrow from the argon excimer emission to the perturbed Wannier-Mott impurity state of xenon-doped liquid argon. The curly vertical arrow denotes the transition which is assumed to be responsible for the newly found near-infrared emission (see, e.g., Fig.\,1, upper panel in ref.\,\cite{neumeier_epl}). After a collisional deexciation of the xenon atoms to the $^{3}P_{1}$ and $^{3}P_{2}$ states, xenon excimers are formed which decay into the dissociative ground state, thereby emitting the characteristic xenon excimer radiation in the vacuum ultraviolet (see Fig.\,3 in ref.\,\cite{neumeier_epl_2}).
    
The concentration-dependent behaviour of the intense near-infrared emission with increasing xenon concentration depicted in Fig.\,2 in ref.\,\cite{neumeier_epl} could be explained by the present model as follows: An increasing concentration of xenon atoms in the Wannier-Mott impurity state $\left(n=2,\,^{2}P_{\frac{3}{2}}\right)$ in liquid argon initially leads to an increasing near-infrared signal. When the xenon-doping concentration becomes so high that the xenon atoms begin to aggregate and form clusters, the excitation energy of the Wannier-Mott impurity state $\left(n=2,\,^{2}P_{\frac{3}{2}}\right)$ begins to rise from the value in the liquid phase ($\sim9.8$\,eV) to the value in the solid phase ($\sim10.1$\,eV \cite{jortner}) and is therefore shifted to a region where the argon excimer emission becomes weaker. This could lead to a decreasing energy transfer and, thus, to a decreasing near-infrared signal with increasing xenon concentration. With increasing concentration there is a strong trend of xenon atoms to form clusters due to the low temperature of liquid argon. Actually, for xenon concentrations larger than 30\,ppm in pure liquid argon a xenon-aggregation has been observed \cite{jortner}. This concentration is in agreement with the results obtained in ref.\,\cite{neumeier_epl} where the peak near-infrared emission has been observed with a concentration of 10\,ppm xenon in liquid argon. 
    
Furthermore, the measured time structure of the near-infrared emission and the following xenon excimer emission (see Fig.\,6 in ref.\,\cite{neumeier_epl_2}) can also be explained qualitatively since the xenon excimer emission in the vacuum ultraviolet is a transition in a cascade following the near-infrared emission. 

\section{Summary}

Due to the improved distillation procedure no attenuation of VUV light in pure liquid argon could be measured. Therefore, no final value for the attenuation length has been stated but rather a lower limit of 1.10\,m. This demonstrates that pure liquid argon has a larger attenuation length than measured in a previous publication \cite{neumeier_epjc}. From the present study it has to be noted that wavelength-resolved attenuation-length measurements are basically absolute optical measurements which are difficult to handle although the concept is straightforward. This is especially the case for the measurements presented here in the vacuum ultraviolet wavelength region. To give a reliable statement on such an absolute value, every unknown parameter has to be determined exactly (e.g., wavelength-dependent refractive indices of liquid argon and of magnesium fluoride in the VUV, see also the discussion on the "fogging effect" in subsection 2.2 in ref.\,\cite{neumeier_epjc}) prior to the measurements presented here. These aspects will be pointed out in a very detailed way in a separate publication focusing on the technical aspects of such measurements \cite{neumeier_jinst}. 

One major result in the context of particle detectors could be confirmed and has to be emphasized: Xenon can act as a critical impurity in pure liquid argon scintillation detectors. For a concentration above 0.1\,ppm xenon in liquid argon, the mixture is opaque for the scintillation light of pure liquid argon below $\sim130$\,nm (see for comparison the emission of pure liquid argon in Fig.\ref{fig:lar_emission_transmission}, lower panel) for an optical path length of 11.6\,cm. Therefore, an additional purification procedure besides conventional chemical purification is mandatory (e.g., fractional distillation) since xenon concentrations of 0.1 to 1\,ppm can be expected from commercial sources of argon gas. A calibration curve connecting the area of the 140\,nm transmission minimum with the xenon concentration is presented in Fig.\,\ref{fig:wachstumskurve} and can be used to estimate xenon concentrations in other liquid noble gas samples. Finally a model qualitatively explaining the near-infrared emitting species discovered in ref.\,\cite{neumeier_epl} is presented and discussed.

\section{Outlook}

To provide an improved value for the attenuation length of pure liquid argon a longer transmission cell is needed to increase the sensitivity. Furthermore, the wavelength-dependent refractive index of pure liquid argon has to be measured precisely. A further improvement would be a light source with a stronger emission in the short-wavelength region \cite{dandl_epl} to explore the region where pure liquid argon becomes intrinsically intransparent. A light source which has a smoother emission shape than a deuterium lamp would be preferred due to the discrete lines in the deuterium spectrum which lead to technical problems when measurement and reference spectrum are not perfectly aligned. A monochromator in front of the inner cell would eliminate the problem of resonance fluorescence (see Fig.\,\ref{fig:transmission_larxe}). It should also be noted here that a setup where the cell is filled vertically instead of horizontally would lead to the possibility of measuring different optical path lengths in the liquid just by adjusting the filling level. This could reduce several systematic errors since the relative change from one level to another can be measured without changing the setup.

A further improvement would be a more accurate measurement of the xenon concentration in the liquid phase of the noble gas samples. One could think of taking samples from the liquid phase and measuring the transmission of the evaporated sample in the gas phase. A comparison of the transmission spectrum of the evaporated liquid mixture with a spectrum of the gas phase with preciseley known xenon concentrations would then provide information on the concentration in the liquid phase.  

\section*{Acknowledgements}
This research was supported by the DFG cluster of excellence "Origin and Structure of the Universe" (www.universe-cluster.de) and by the Maier-Leibnitz-Laboratorium in Garching.

\end{document}